# Controllable selective exfoliation of high-quality graphene nanosheets and nanodots by ionic liquid assisted grinding†

Nai Gui Shang,*[a] Pagona Papakonstantinou,*[a] Surbhi Sharma,[a] Gennady Lubarsky,[a] Meixian Li,[b] David W. McNeill,[c] Aidan J. Quinn,[d] Wuzong Zhou[e] and Ross Blackley[e]

**Bulk quantities of graphene nanosheets and nanodots have been selectively fabricated by mechanical grinding exfoliation of natural graphite in a small quantity of ionic liquids. The resulting graphene sheets and dots are solvent free with low levels of naturally absorbed oxygen, inherited from the starting graphite. The sheets are only two to five layers thick. The graphene nanodots have diameters in the range of 9–29 nm and heights in the range of 1–16 nm, which can be controlled by changing the processing time.**

Recently graphene has received extensive attention since it has demonstrated many unique electrical, thermal and mechanical properties that have never been found in other materials.[1–5] It has been both theoretically predicted and experimentally proved that the size, composition and edge geometry of graphene are important factors, which determine its overall electronic, magnetic, optical and catalytic properties due to strong quantum confinement and edge effects. For example, by cutting graphene sheets into long and narrow ribbons (GNRs) (width less than 10 nm) it is possible to induce a direct band gap in graphene, which renders GNRs semiconducting.[6] Further confinement in the basal plane (overall dimensions smaller than 100 nm) leads to quantum dots (GQDs) with zero dimensions. The suppressed hyperfine interaction and weak spin–orbit coupling make GQDs interesting candidates for spin qubits with long coherence times for future quantum information technology.[7] Therefore graphene sheets with reduced lateral dimensions in the form of nano-ribbons or quantum dots can effectively tune the band gap of graphene and facilitate the lateral scaling of graphene in nanoelectronic devices. In this context it has become urgent to develop effective routes for tailoring the graphene structures.[8,9]

To date, three main methods such as chemical vapour deposition (CVD),[10] micromechanical cleavage and chemical exfoliation[11] have been used to fabricate graphene sheets. Compared to other techniques, chemical exfoliation, which involves the direct exfoliation of various solid starting materials, such as graphite oxide, expanded graphite and natural graphite,[12–15] is advantageous in terms of simplicity, low cost and high volume production. However, currently explored chemical solution exfoliation methods have three main drawbacks that need to be addressed. Firstly the produced graphene is quite poor in quality compared to that fabricated by CVD and micromechanical cleavage. This is mainly because the various chemicals used, such as solvents, oxidants and reductants, may attack the graphene lattice in the process or are difficult to be removed, leading inevitably to residual surface species. Overall these chemical processes introduce various forms of surface defects, which disrupt the graphene band structure and hamper the conductivity of the resulting graphene sheets. Secondly, many of the chemicals used are either expensive or toxic and need careful handling,[16] leading to environmentally unfriendly and unsustainable practices. Thirdly, the majority of chemical solution exfoliation methods involve extremely time-consuming multiple steps that sometimes last for several days or weeks. Therefore, in order to overcome the above-mentioned limitations and obtain high-quality graphene, it is necessary to develop simple, rapid chemical exfoliation methods which utilise cheaper and more "environmentally friendly" chemicals. To date, some progress has been achieved. For example, Wang *et al.* have reported that few layer graphene sheets can be directly exfoliated from natural graphite by using tip ultrasonication in ionic liquids.[17] The use of natural graphite may not only decrease the cost compared to that of the expanded graphite or graphene oxide, but also can improve the quality of the resulting graphene due to the absence of oxygen-containing groups. However, the graphene sheets produced by these simple techniques still contain a few impurities (F and S *etc.*), and a large fraction of oxygen[17] (>10 at%) similar to those found in graphene reduced from graphene oxide. Oxygen in graphene is difficult to be removed[18] and may significantly influence its properties. Therefore, it is highly desirable to develop new high-yield methods to make high-quality graphene sheets.

[a] *Nanotechnology and Integrated Bio-Engineering Centre, NIBEC, School of Engineering, University of Ulster, Newtownabbey, BT37 0QB, UK. E-mail: p.papakonstantinou@ulster.ac.uk, ngshang@hotmail.com*
[b] *College of Chemistry and Molecular Engineering, Peking University, Beijing, 100871, P.R. China*
[c] *School of Electrical Engineering, Queen's University of Belfast, Belfast, BT9 5AH, UK*
[d] *Tyndall National Institute, University College Cork, Lee Maltings, Dyke Parade, Cork, Ireland*
[e] *School of Chemistry, University of St Andrews, St Andrews, KY16 9ST, Scotland, UK*

† Electronic supplementary information (ESI) available: Details of the fabrication process, narrow scan XPS, sheet and nanodots distribution, additional TEM images, height profile of nanodots, Raman scattering, thermal gravimetric analysis, optical property of nanodots, electrical properties of pristine and vacuum annealed graphene sheets studied by a four-probe technique. See DOI: 10.1039/c2cc17185f

In this study, we report the direct exfoliation of natural graphite into high-purity few layer graphene sheets and nanodots by using a novel environmentally friendly method, involving simple ionic liquid (IL) assisted grinding to produce a gel, followed by a cleaning step to remove the IL. Ionic liquids are chosen because they are green organic solvents with a surface tension well matching the surface energy of graphite, preventing the detached graphene from restacking.[19,20] It should be emphasized that our procedure is different from other reported studies, where either prolonged or high intensity ultrasound is the driving force for the exfoliation. Our procedure is mild and relies on pure shear forces to detach the graphene layers from the graphite flakes. Therefore, severe defect formation on the crystalline plane of graphene, or chemical reactions due to cavitation effects induced during sonolysis are avoided resulting in high quality material. The process used is described in detail in the ESI.†

Fig. 1(a) and (b) show typical XPS survey scan spectra of graphene products and the starting powder of graphite flakes, respectively. They both show a strong C1s peak at 284.5 eV, a small O1s peak at 532.6 eV and a weak OKLL Auger band between 955–985 eV. Except for oxygen and carbon, no other elements such as F, N or P from the chemicals used (IL and N,N-dimethylformamide) are found in the sample. The concentration of elements C and O in graphene is calculated to be about 96.4 and 3.6 at%, respectively, very close to those in the starting graphite powder (3.4 at% of O). This demonstrates that the graphene sheets are clean and free of any impurities and contaminations from the chemicals used, except for a small amount of oxygen inherited from the starting graphite material. This is in stark contrast to graphene sheets produced by tip ultrasonication in ionic liquids,[17] where impurities (F and S etc.) inherited by the IL, and a large fraction of oxygen (more than 10 at%) are present.

Depending on the preparation parameters, two kinds of graphene structures can be formed in the supernatant (20 wt% of the starting material) of the sedimentation process: submicron width few-layer sheets and nanometre-sized nanodots. Submicrometre graphene sheets are dominant in the supernatant, when a grinding time of less than 30 min and a ratio of graphite flakes (mg) to ionic liquid (μL) of 1:10 up to 1:4 are applied. Fig. 2(a) shows a typical TEM image of a collection of graphenes. Additional TEM images and SEM images of graphene sheets are available in Fig. S2 and S5 (ESI†). It is found that graphene sheets have a size of 0.006–0.36 μm$^2$ and some are stacked together. Fig. 2(c) shows a size distribution of a total of 93 distinguishable graphene sheets calculated from TEM images using software ImageJ. Sheet sizes of 0.006–0.0125 μm$^2$ are dominant representing 50% of the total distribution. Around 22% of sheets have a size of 0.022–0.038 μm$^2$, and only a few larger sheets with a size up to 0.3 μm$^2$ are present. High resolution TEM analysis of the graphene edges reveals that the majority of the graphene sheets are made of 2–5 layers, with a lattice spacing of 0.342 nm (see Fig. 2(d)–(f)). No other carbon phases such as amorphous carbon or fullerene etc. are found at the edges. The corresponding electron diffraction pattern of single sheets (Fig. 2(b)) has a typical six-fold symmetry, confirming that the graphene sheet is of high-quality single crystal nature. When a longer grinding time and a smaller quantity of ionic

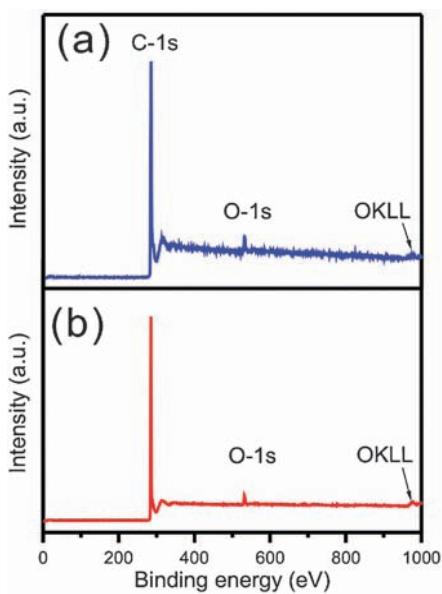

**Fig. 1** (a) and (b) XPS survey scan spectra of graphene sheets and starting graphite, respectively.

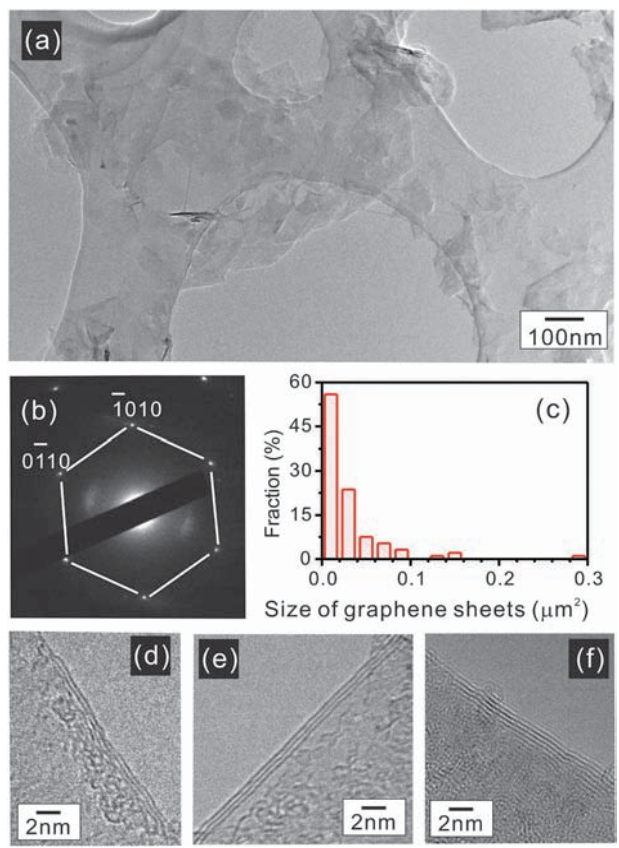

**Fig. 2** (a) A typical low-magnification TEM image of graphene sheets; (b) corresponding electron diffraction pattern of (a); (c) size distribution of graphene sheets; HRTEM images of (d) bilayer, (e) triple layer and (f) 4–5 layer graphene sheets.

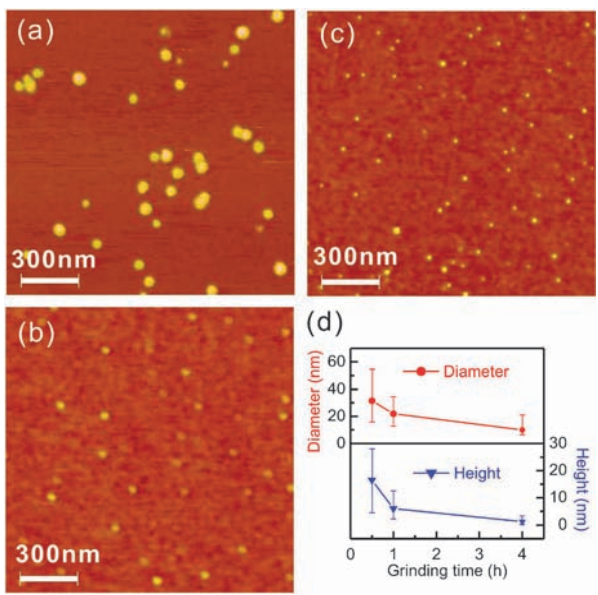

**Fig. 3** AFM images of nanodots, produced with grinding times of (a) 30 min, (b) 1 h and (c) 4 h and a ratio of 1 : 2 of graphite (mg) to ionic liquid (mL); (d) height and diameter distributions of nanodots.

liquid are applied (1 : 2 ratio), graphene nanodots are dominant in the supernatant of the centrifugation process. The diameter of graphene nanodots can be controlled by changing the grinding time. Fig. 3(a)–(c) show typical AFM images of graphene nanodots, which are produced with a grinding time of 0.5, 1 and 4 hours, respectively.

The nanodots are dispersed on Si or freshly cleaved mica surfaces and measured by an AFM operating in the tapping mode. The nanodots have an average diameter of 29 nm for the grinding time of 30 min. When the grinding time is increased to 1 and 4 h, the average diameter of the nanodots reduces to 20 and 9 nm, respectively. Besides the diameter, the height of the nanodots also decreases as the grinding times increases (Fig. 3(d)). For 30 min grinding the average height of the nanodots is approximately 16 nm. This reduces dramatically to 5 and 1 nm, respectively, for grinding times of 1 and 4 h (Fig. S4, ESI†). Measured lateral distributions are provided in Fig. S3 (ESI†). Meanwhile, the nanodots produced for different grinding times have quite different lateral and height distributions. Longer grinding times yield narrower distributions. For a grinding time of 4 h, the nanodots present a monodispersion in height of only 1–3 nm (few layer graphene). HRTEM images and optical properties of graphene nanodots are provided in Fig. S9 and S10 (ESI†). In addition to microscopic characterization of graphene, macroscopic techniques such as Raman scattering and thermal gravimetric analysis of bulk quantities have been used to characterize our samples as well (Fig. S6 and S7, ESI†). Both techniques show the graphene sheets and nanodots are of high quality.

In summary, we have developed a new controllable method to selectively produce few-layer graphene sheets and nanodots with a high yield ($\sim$20%) from natural graphite by using a simple grinding method with ionic liquid as the grinding agent. The produced graphene sheets are free from chemical functionalities and consist of high-quality single crystals with only two to five layers. The graphene nanodots have a diameter ranging from 9 to 29 nm and a height ranging from 1 to 16 nm, which strongly depend on the grinding time. The formation of high-quality graphene is achieved by a green procedure different from other reported studies, where either prolonged or harsh sonication is the driving force for the exfoliation of graphite flakes. The present method has the potential to exfoliate other layered materials such as $MoS_2$ or BN in addition to graphite.

The work was supported from INVEST Northern Ireland (Proof of Concept Award POC114), INVEST NI and the University of Ulster (Joint Proof of Concept Award POC114), Royal Academy of Engineering/Leverhulme Trust Senior Research Fellowship (to PP), EPSRC funded facility access to HRTEM at the Un. of Nottingham and Un. of St Andrews, and the Tyndall National Access Programme, NAP supported by SFI. J. Benson's help with UV-Vis is acknowledged.

Bulk quantities of graphene sheets and nanodots have been selectively fabricated by mechanical grinding exfoliation from natural graphite in a small quantity of ionic liquid.

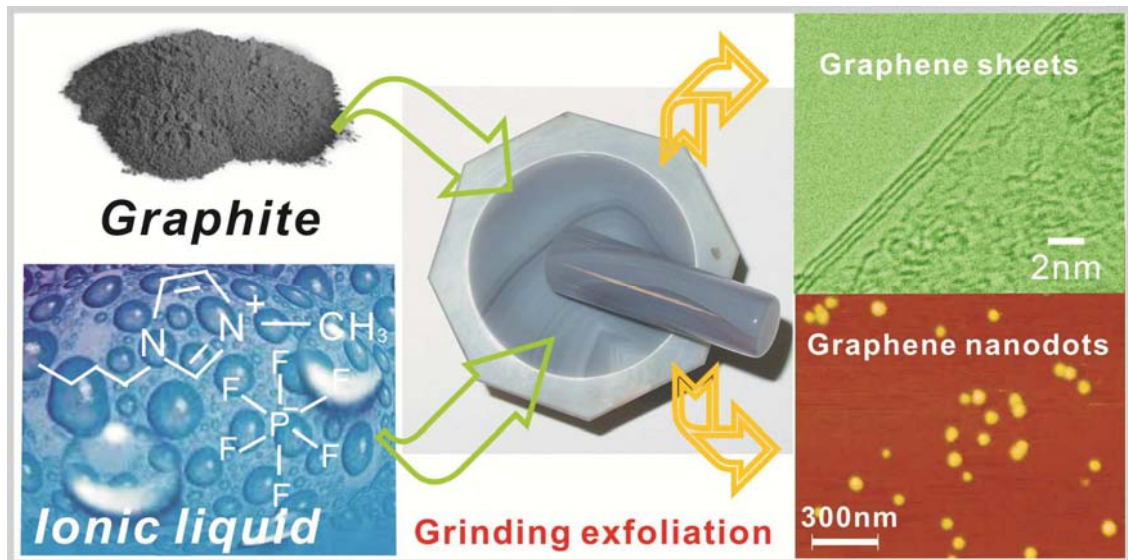

# Supporting information

# Controllable selective exfoliation of high-quality graphene nanosheets and nanodots by ionic liquid assisted grinding

Nai Gui Shang[*], Pagona Papakonstantinou[*], Shurbhi Sharma[†], Genaddy Lubarsky

Nanotechnology and Integrated Bio-Engineering Centre, NIBEC, School of Engineering, University of Ulster, Newtownabbey, BT37 0QB, UK

Meixian Li,

College of Chemistry and Molecular Engineering, Peking University, Beijing, 100871 P.R. China

David W. McNeill

School of Electrical Engineering, Queen's University of Belfast, Belfast, BT9 5AH, UK

Aidan. J. Quinn

Tyndall National Institute, University College Cork, Lee Maltings, Dyke Parade, Cork, Ireland

Wuzong Zhou and Ross Blackley

School of Chemistry, University of St Andrews, St Andrews, KY16 9ST, UK

*corresponding authors: p.papakonstantinou@ulster.ac.uk; ngshang@hotmail.com

[†] Current address: School of Chemical Engineering, The University of Birmingham Edgbaston, Birmingham B15 2TT, U.K

## 1. Fabrication process

Graphite powders of 50 mg (99.0% purity, from Sigma-Aldrich) with a grain size less than 20 μm were mixed and ground with 0.1—0.5 mL ionic liquid (IL, 1-Butyl-3-methylimidazolium hexafluorophosphate, BMIMPF$_6$, from Sigma-Aldrich) for 0.5 — 4 hrs. Then the mixtures were added into a solution of 15 mL N,N-dimethylformamide (DMF) and 15 mL acetone and centrifugated in the speed of 3000 rpm for 30 min in order to remove the ionic liquid. This washing cycle was repeated three times and the final sediment was dispersed in 1 L DMF. After one-day sedimentation, the large/thick graphitic flakes completely precipitated on the bottom of the bottle. The supernatant was dried and then the yield was calculated as 20 wt%, which is much larger than that of other methods reported. The supernatant was collected and centrifugated to get a dense suspension, with which the graphene samples were prepared for the study of their microstructure and electrical properties. The suspension was diluted and dropped on lacy carbon-coated Cu grid for TEM observations, on clean Si wafers or mica sheets for the XPS, Raman and AFM studies, and on thermally oxidised Si wafer substrates (300 nm SiO$_2$) for the study of electrical properties. Note, to obtain uniform large-area graphene thin films, the Si/SiO$_2$ substrate was functionalised with 5% 3-triethoxysilylpropylamine (APTES) aqueous solution for 30 min. All samples were completely dried under an infrared light or on a hot plate before measurements.

We should note the role of ILs in this study. ILs are low-temperature molten salts, and as their name denotes are liquids composed entirely of ions. They have been proposed as a new class of "green" organic solvents because not only they are able to solvate a large variety of organic and inorganic compounds, (polar or non-polar) but also they possess high thermal and chemical stability, high ionic conductivity, wide electrochemical window, and negligible vapour pressure. The mixing of a small quantity of IL and graphite flakes forms a gel-like composite. When graphite flakes are ground with ionic liquids, the shear force detaches the graphene layers from the graphite flakes. The ionic liquids can effectively surround each layer preventing the detached graphenes from restacking. Based on simulation and experimental studies on processing of carbon nanotubes with imidazolium-based ILs, the electronic structure of graphene layers in the bucky gel remains unchanged and there is no charge transfer between graphene and imidazolium cations[1].

2. XPS data analysis

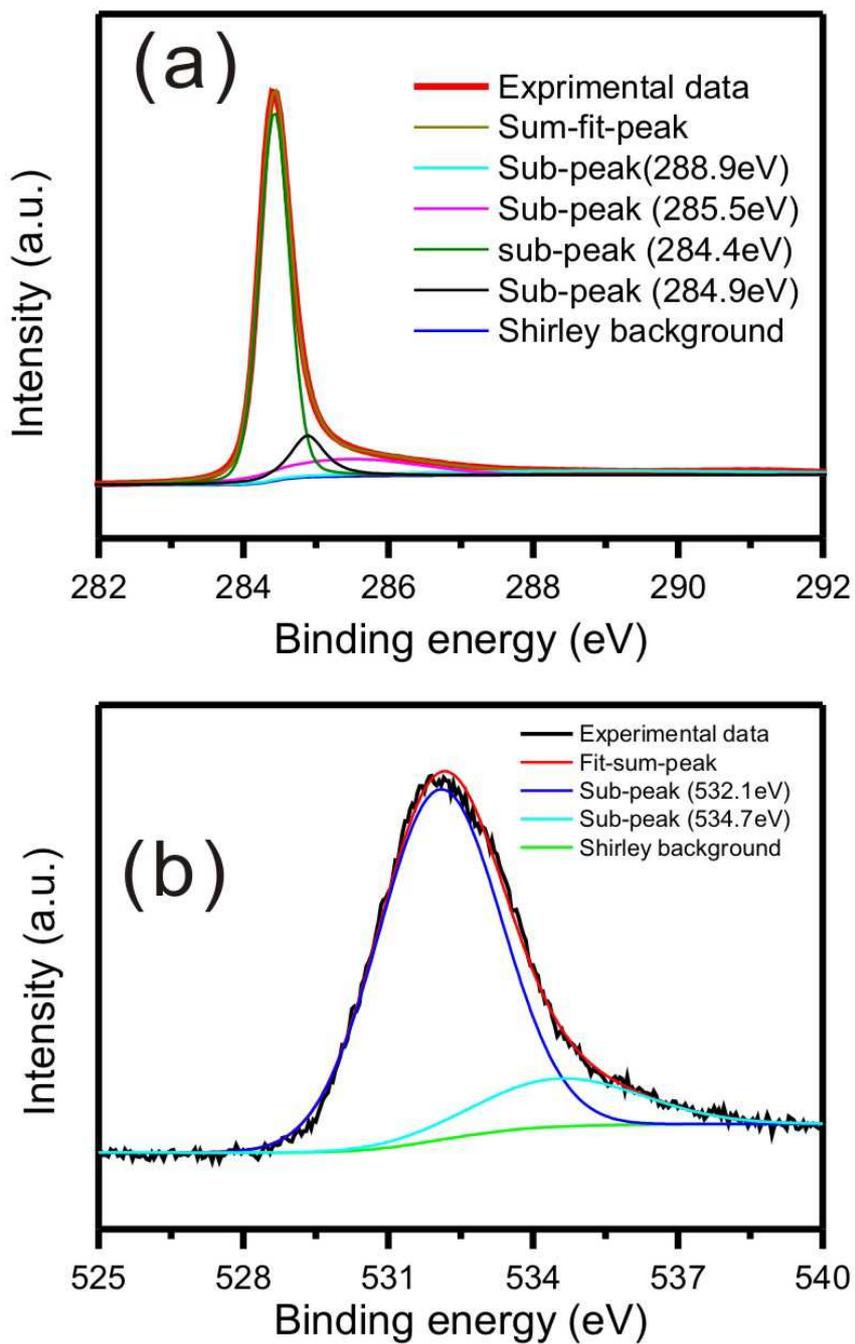

**Figure S1.** XPS narrow scan spectra of graphene samples for elements C 1s (a) and O1s (b), respectively.

Figure S1 (a) and (b) present a high resolution asymmetric C1s and an O1s XPS spectrum of graphene sheets, respectively. After subtraction of a Shirley background followed by a fitting process using a mixture of Lorentzian and Gaussian lineshapes, the C1s peak was deconvoluted into four sub-peaks located at 284.4, 284.9, 285.5, and 288.9 eV, which have been assigned to C-C ($sp^2$), "defect peak", C-O and COOC/COOH bonds, respectively. The O1s peak can be fitted by two Gaussian peaks at 532.1 and 534.7 eV corresponding to C-O and C=O bonds.

3. Low magnification TEM images of graphene nanosheets and their size distributions

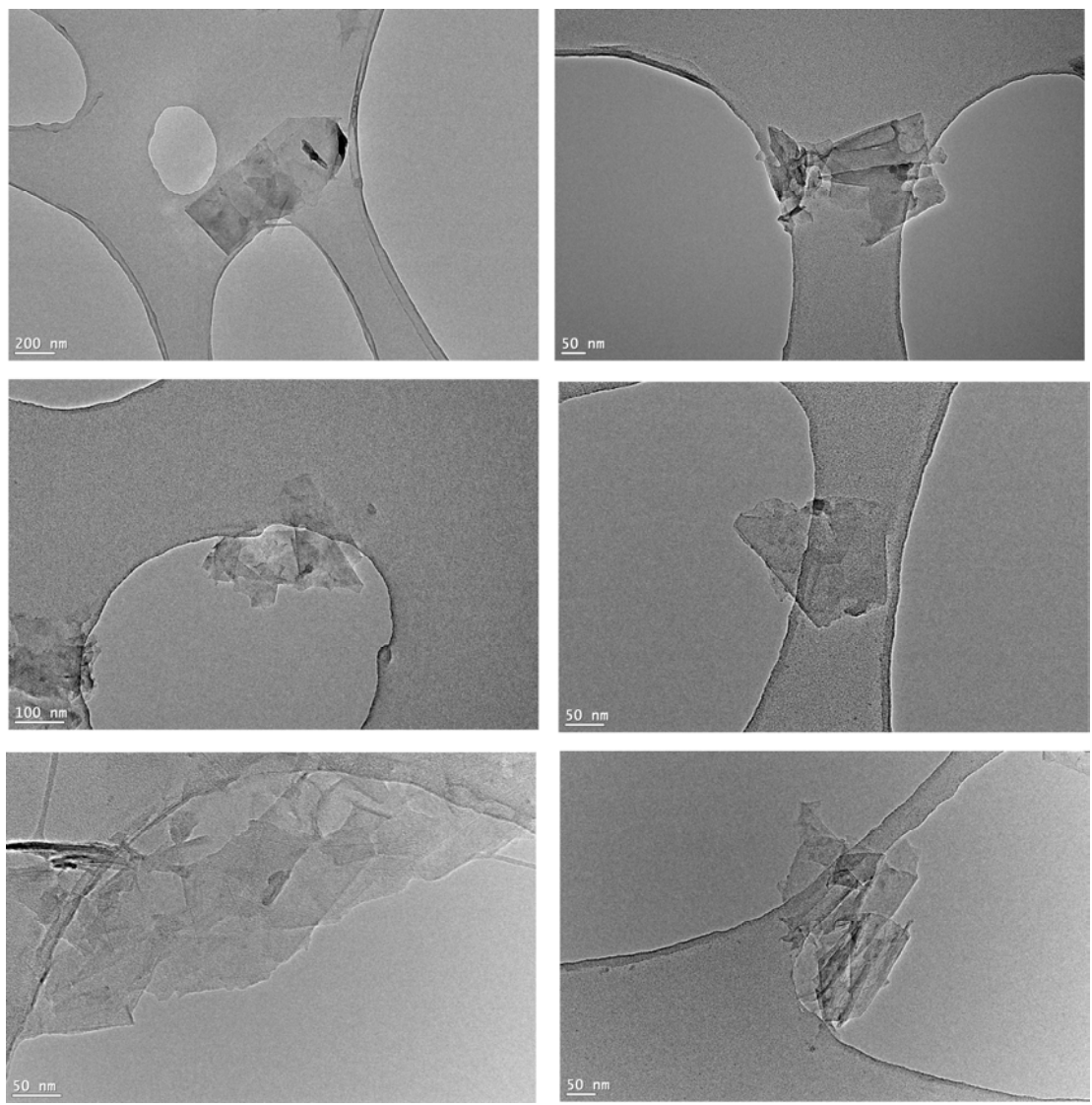

**Figure S2.** Low magnification TEM images of folded and wrinkled graphene nanosheets.

## 4. Size distribution of graphene nanodots

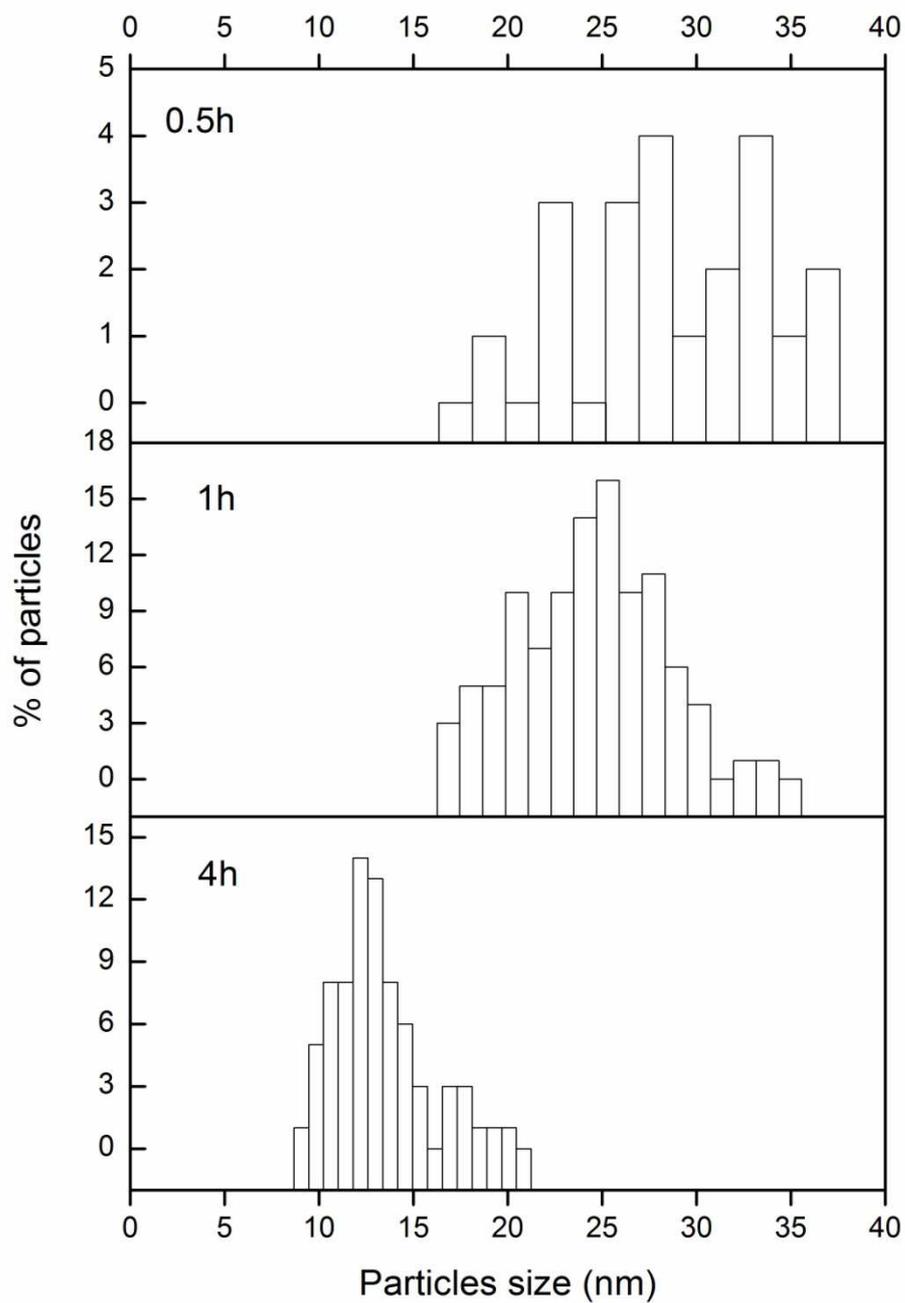

**Figure S3.** Histogram of graphene nanodots describing their size distribution.

## 5. AFM images and profile line scans of graphene nanodots

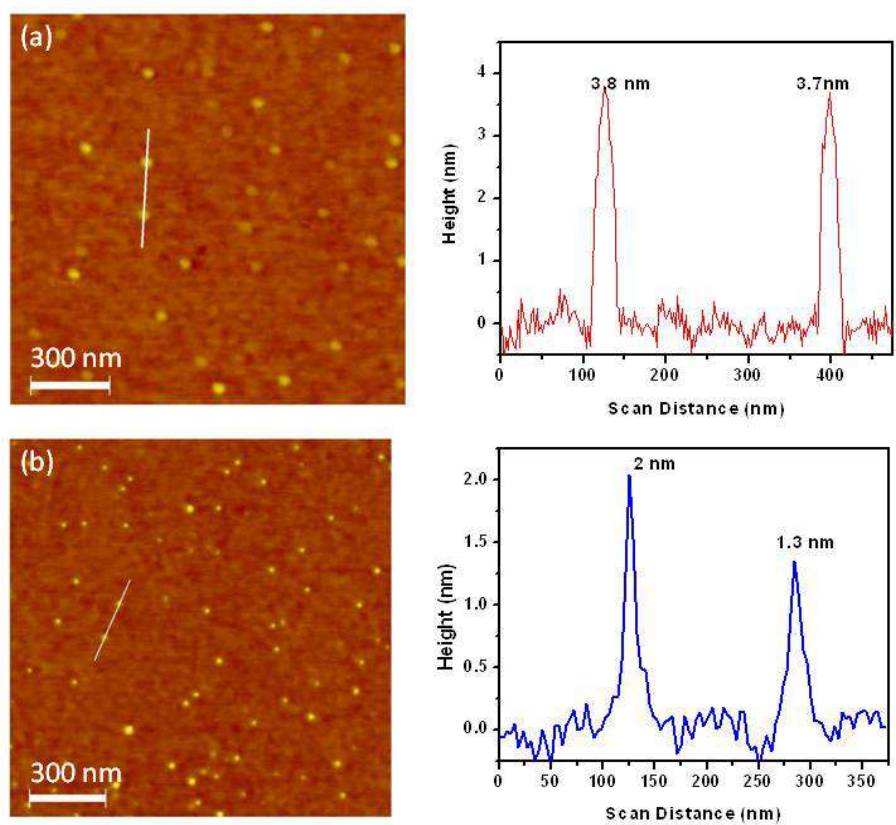

**Figure S4.** Typical Atomic Force Microscopy images and topography line profiles for graphene nanodots produced using different grinding times: (a) 1 hour, (b) 4 hours.

## 6. SEM images of graphene nanosheets

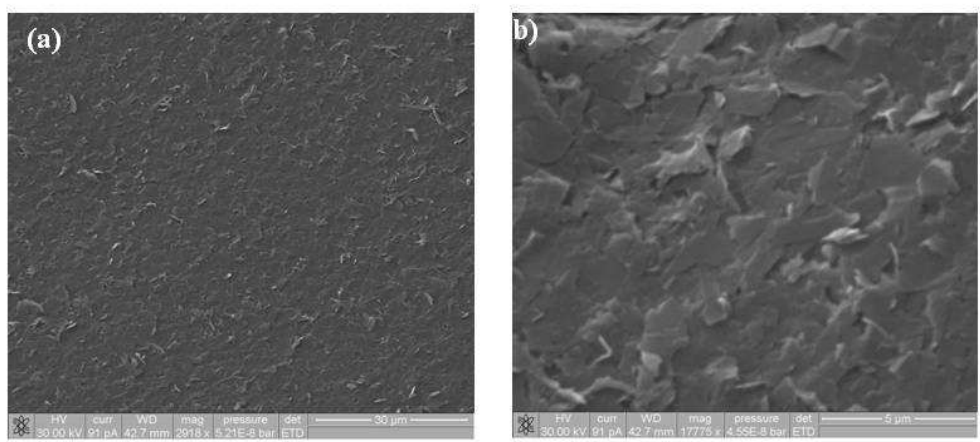

**Figure S5.** Different magnification SEM images of graphene sheet films produced through filtration of graphene solution.

## 7. Raman spectrum of graphene sheets and nanodots

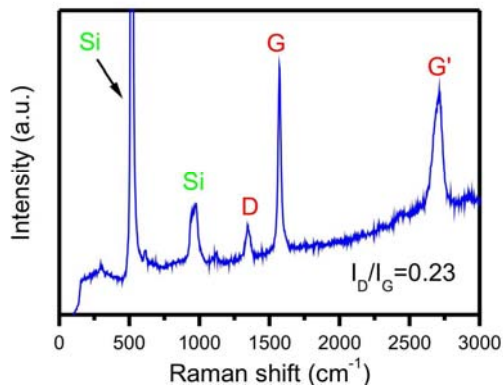

**Figure S6.** Typical Raman spectrum of the graphene nanosheets.

Raman scattering is a convenient, powerful macroscopic tool for the characterisation of graphene. The layer number and quality of graphene fabricated by the method of micromechanical cleavage can be well distinguished by the analysis of spectroscopic intensity, frequency and line width etc. Figure S6 shows a typical Raman spectrum of the graphene sample. The Raman measurement is conducted in a backscattering geometry at room temperature using an excitation laser of 514 nm with a spot size of 1-2 μm. There are three strong peaks at 1348, 1571, and 2711 cm$^{-1}$, which are ascribed to the D, G, and G´ bands of graphitic materials, respectively. No broad peaks relevant to amorphous carbon are found in the sample. The presence of the D peak arises from the edges of graphene sheets, whoose size is smaller than the laser spot (1-2μm). The ratio of integrated intensities of D to G bands ($I_D/I_G$) is only 0.23, revealing that the graphenes we produced are of high quality, since the D band is a fingerprint of defects in graphitic materials. The G´ band, a fingerprint of graphene, is quite strong, larger than the D but slightly smaller than the G band in intensity. The ratio of integrated intensity of G´ to G bands is about 0.61, revealing the presence of 2-4 graphene layers consistent with the TEM observation. Note, currently the Raman spectra of graphene fabricated by solution based methods do not resemble those of graphene produced by other methods. They do not present identical spectroscopic features and are strongly dependent on the fabrication method and chemicals used. Most of them present a strong D band, a broad G band and a weak G´ band. However, all findings presented here confirm that the crystalline quality of graphene we produced is better to those reported by solution methods in other groups, but poorer than the graphene produced by mechanical cleavage of highly oriented pyrolytic graphite and by high-temperature CVD growth on metal substrates, where no defect related D bands can be detected. The relatively poorer quality of graphene sheets is believed to be due to the low-quality starting material (natural graphite) compared to the HOPG, and not due to the grinding process.

## 8. TGA spectra of the starting graphite and produced graphene

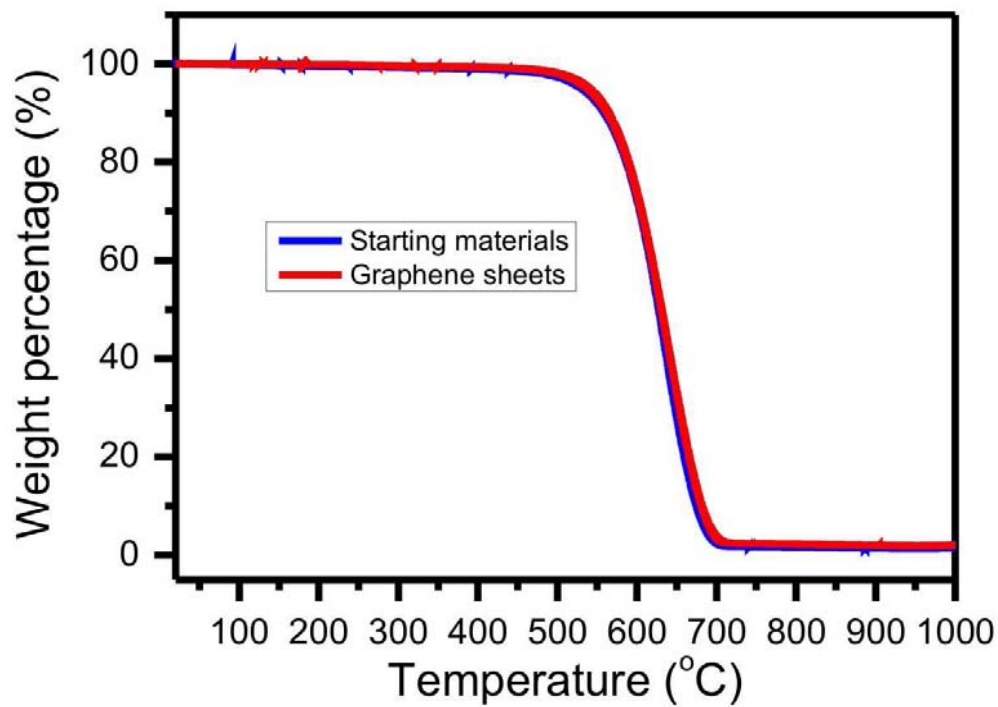

**Figure S7.** TGA spectra of graphene sheets and starting graphite flakes material.

Figure S7 shows TGA spectra taken from the graphene sheets and the starting graphite flakes material. The measurement was done in a mixture ambient of 50 % $N_2$ and 50 % $O_2$ with a ramp of 1 °C/min. It can be seen that graphene sheets have a same thermal behaviour as that of the starting material. They simultaneously start to oxidize at around 500 °C and completely burn off at 700 °C, demonstrating exactly same quality and no any high/low melting materials being introduced in the fabrication process.

## 9. Electrical resistivity of pristine and vacuum-annealed graphene nanosheet films

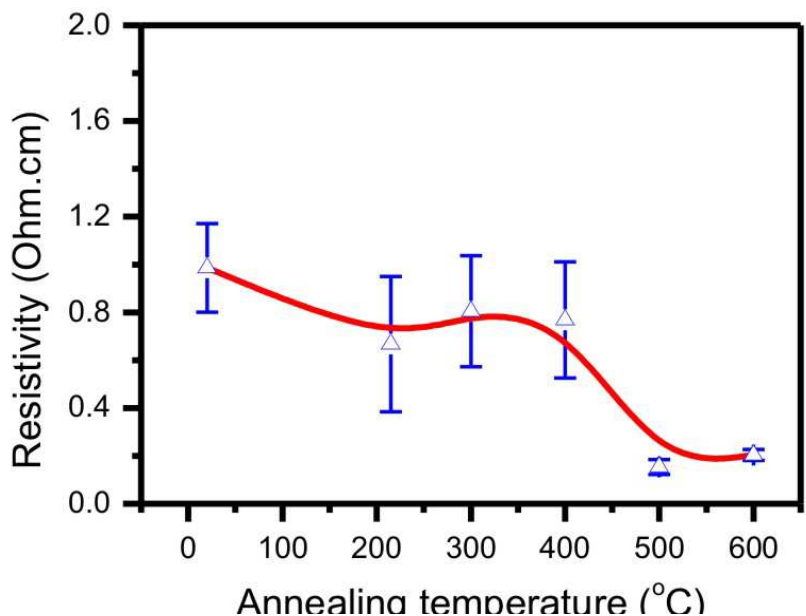

**Figure S8.** The resistivity of graphene sheet films as a function of the annealing temperature.

Figure S8 shows the resistivity of graphene films as a function of the annealing temperature. The graphene films with a thickness of around 12.8 μm were fabricated on a large SiO$_2$ coated Si wafer by drop coating and then were cut into several 10×20 mm$^2$ sized pieces for annealing in vacuum (less than 8×10$^{-5}$ Torr) for 1 hr in the temperature range of 200 - 600 ºC. The resistances were measured by a four-probe technique at room temperature (Resistivity test rig, Model B, A & M Fell LTD, England). The resistivity of pristine graphene films is about 1×10$^{-2}$ Ω•m. The resistivity decreases to 8×10$^{-3}$ Ω•m with the increase of the annealing temperature in the range of 200 - 400 ºC. When the annealing temperature is increased to 500 - 600 ºC, their resistivity significantly lowers to 2×10$^{-3}$ Ω•m. However, this value is still larger than that (0.07 - 0.11×10$^{-3}$ Ω•m) of graphene films produced by other chemical solution methods.[2] The main reason is that there are a lot of junctions in the film due to either the smaller size of single graphene sheets or the presence of voids. The standard deviations of pristine and annealed graphene films at low temperatures are quite large, suggesting that the graphene film could have different thicknesses and different numbers of internal structural voids. The variation of the graphene resistivity appears to take place in two steps, suggesting that the graphene film is subjected to two kinds of transitions during annealing. The first process could be due to the desorption of various absorbates such as water, C-H and COOH groups from the surface of graphene films and the inner surface of the structural voids. The second step could be ascribed to the shrinking or collapsing of the voids, leading to relocation of all constituent graphene sheets and the formation of a dense uniform film at high temperature, which is evidenced by the small value of standard deviation in the resistivity data at this temperature.

**TEM images of graphene nanodots**

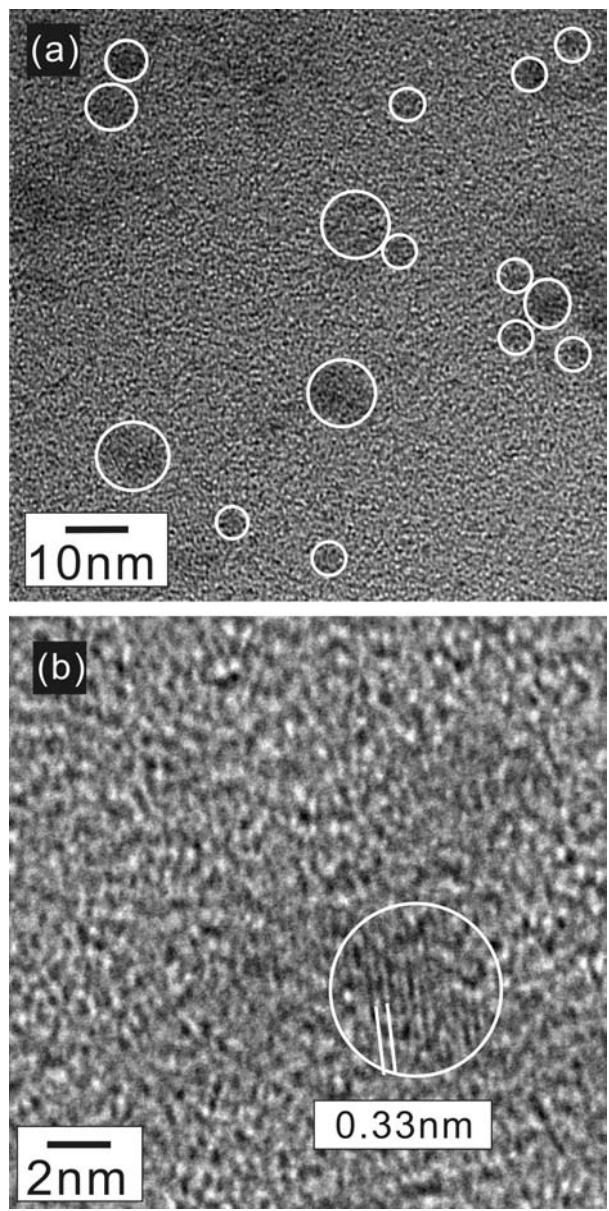

**Figure S9.** Typical TEM images of graphene nanodots.

Figure S9(a) shows a typical low-magnification TEM images of graphene nanodots which are produced by grinding for 4hrs. The graphene nanodots have a size of 5-12 nm, in close agreement with the AFM results. The high-resolution TEM image (Fig. S9b) shows the high crystalline quality of nanodots with a lattice spacing of around 0.33nm.

## 10. Optical properties of graphene nanodots

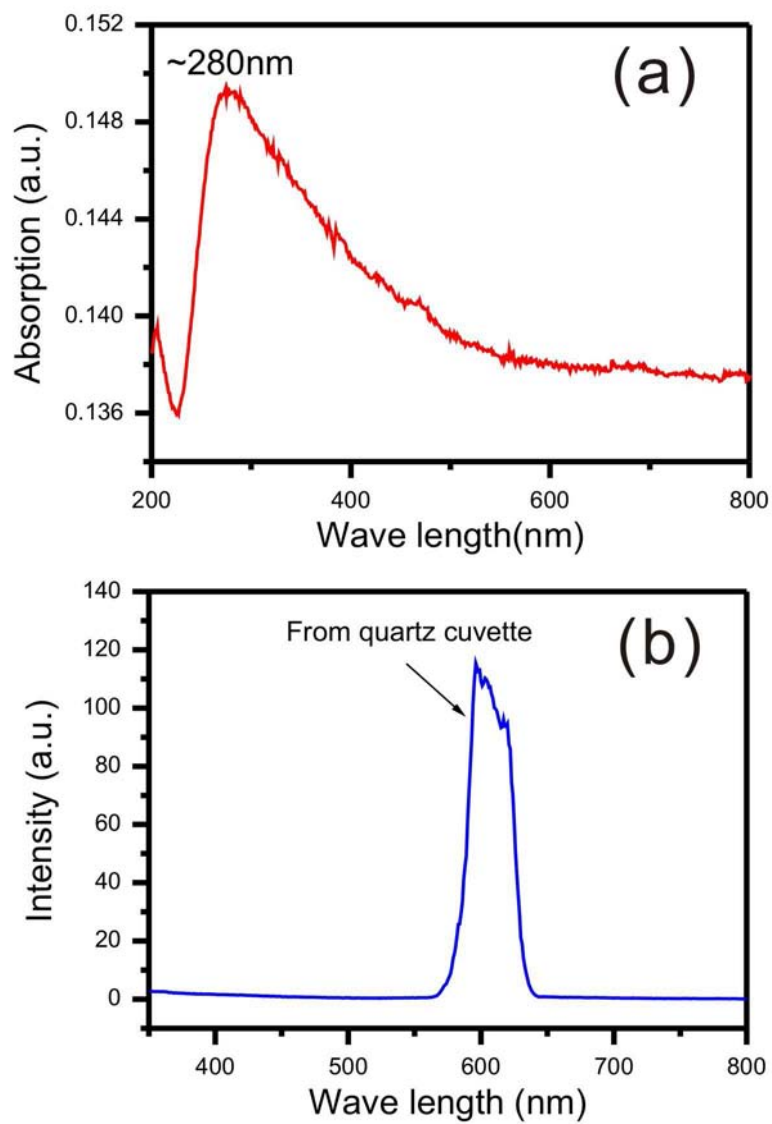

**Figure S10**. (a) UV-Vis absorption and photoluminescence spectra of graphene nanodots.

Figure S10 (a) shows a typical UV-Vis absorption spectrum taken from ~0.1mg/mL graphene nanodots in de-ionized water by using a UV-Vis spectrometer (Perkin Elmer Lambda 35). The graphene nanodots have a broad absorption band centered at around 280nm. Figure S10 (b) shows a typical photoluminescence spectrum taken from ~0.03mg/mL graphene in DMF solution using a Cary Eclipse fluorescence meter with a 330 nm filtered excitation at room temperature in a wavelength range of 350-800nm. It can be seen that there is a strong broad red band centered at 610 nm, which originates from the quartz cuvette rather than from the graphene nanodots. No photoluminescence from the graphene nanodots was observed. This was confirmed by a control experiment with a bare DMF solution. The main reason for the absence of PL in the graphene nanodots produced by IL assisted grinding in the present work is that they are free of both the defects and functional groups, which are usually considered to result in PL.[3][4]